\documentclass[%
reprint,
 amsmath,amssymb,
 aps,
]{revtex4-1}

\usepackage{epsfig}
\usepackage{dsfont}

\usepackage{graphicx}
\usepackage{dcolumn}
\usepackage{bm}

\newcommand{\beq}{\begin{equation}}
\newcommand{\eeq}{\end{equation}}
\newcommand{\bea}{\begin{eqnarray}}
\newcommand{\eea}{\end{eqnarray}}
\newcommand{\nn}{\nonumber \\}

\newcommand\eqn[1]{(\ref{#1})}      
\newcommand\Eqn[1]{Eq.~(\ref{#1})}  
\newcommand\Fig[1]{Fig.~\ref{#1}}  

\newcommand{\C}{\mathcal{C}}
\newcommand{\p}{\partial}

\begin{document}


\title{Langevin description of nonequilibrium quantum fields}

\author{F. Gautier}
 \email{fgautier@apc.univ-paris7.fr}
\author{J. Serreau}%
 \email{serreau@apc.univ-paris7.fr}
\affiliation{%
 Astro-Particle and Cosmology (APC), University Paris 7 - Denis Diderot\\ 10, rue Alice Domon et L\'eonie Duquet, 75205 Paris Cedex 13, France.
}%

\date{\today}

\begin{abstract}

We consider the non-equilibrium dynamics of a real quantum scalar field. We show the formal equivalence of the exact evolution equations for the statistical and spectral two-point functions with a fictitious Langevin process and examine the conditions under which a local Markovian dynamics is a valid approximation. In quantum field theory, the memory kernel and the noise correlator typically exhibit long time power laws and are thus highly non-local, thereby questioning the possibility of a local description. We show that despite this fact, there is a finite time range during which a local description is accurate. This requires the theory to be (effectively) weakly coupled. We illustrate the use of such a local description for studies of decoherence and entropy production in quantum field theory.

 \end{abstract}

\pacs{Valid PACS appear here}
\maketitle


\section{Introduction}
\label{sec:intro}

Understanding the dynamics of time evolving quantum systems is a key issue in many topical areas of physics, from early-universe cosmology to high-energy nuclear collisions, condensed matter physics or ultracold atomic gases. One important line of investigation in nonequilibrium field theory concerns the issue of first principle calculations of far-from-equilibrium (quantum) dynamics. Major breakthroughs have been achieved in recent years, in particular with the use of two-particle-irreducible functional techniques \cite{Berges:2003pc,Berges:2004yj}, and the field is under active development, thanks to the advent of ever faster computers and of new ideas \cite{Berges:2006xc,Borsanyi:2008eu,Bodet:2011qt}.   

An important byproduct of such investigations is that it allows one to bridge the gap between relatively simple situations where direct calculations from the basic equations of quantum field theory (QFT) can be done and more intricate cases, such as, for instance, situations with expanding and/or inhomogeneous backgrounds \cite{Aarts:2007ye,Tranberg:2008ae,Prokopec:2003pj,Prokopec:2004ic}, or situations involving many types of fields and interactions such as lepto/baryo-genesis scenarios \cite{Prokopec:2003pj,Prokopec:2004ic,Anisimov:2010aq,Anisimov:2010dk,Garny:2010nz,Garbrecht:2011aw}, where such calculations are difficult and where one often has to rely on effective descriptions. The most popular such effective approaches are kinetic descriptions \cite{Arnold:2002zm,Boyanovsky:1999cy,Ivanov:1999tj,Blaizot:2001nr,Prokopec:2003pj,Prokopec:2004ic,Anisimov:2008dz,Garny:2010nz,Garbrecht:2011aw,Drewes:2012qw,Herranen:2008di}, or effective Langevin descriptions \cite{Starobinsky:1994bd,Biro:1997va,Rischke:1998qy,Bodeker:1998hm,Berera:2004kc,Yokoyama:2004pf,Weenink:2011dd,Nahrgang:2011mg}. A key issue in this context concerns the domain of validity of such approaches.

The derivation of kinetic -- e.g. Boltzmann -- equations from nonequilibrium Schwinger-Dyson, or Kadanoff-Baym equations have been extensively discussed since the early works of Kadanoff and Baym \cite{Kadanoff:1994,Danielewicz:1982kk,Calzetta:1986cq,Boyanovsky:1999cy,Ivanov:1999tj,Blaizot:2001nr,Prokopec:2003pj,Herranen:2008di,Herranen:2008hu,Fidler:2011yq}. This typically relies on a gradient expansion, which assumes a clear separation between the scale of the nonequilibrium dynamics at hand and that of the elementary excitations and processes responsible for it. The validity of the gradient expansion and of the corresponding transport equations has been studied against direct solutions of the underlying nonequilibrium Schwinger-Dyson equations in cases where the latter can be done \cite{Juchem:2003bi,Juchem:2004cs,Berges:2005md,Lindner:2005kv,Lindner:2007am,Garbrecht:2011xw}.

Effective Langevin equations and their derivation from basic QFT have also been much studied, mainly along the lines of the pioneering work of Feynman and Vernon, based on the so-called influence functional formalism 
\cite{Feynman:1963fq,Gleiser:1993ea,Greiner:1996dx,Greiner:1998vd,Rischke:1998qy,Berera:2004kc,Boyanovsky:2004dj}. The latter provides a systematic way of integrating out irrelevant degrees of freedom in a path integral language. One typically obtain non-local -- i.e. non-Markovian -- Langevin-like equations with various additive and/or multiplicative random noises \cite{Gleiser:1993ea,CassolSeewald:2007ru}. Noise correlators and memory kernels can in principle be computed by means of standard QFT techniques. 

The most popular use of Langevin descriptions involves further drastic simplifications, namely the assumption of effective local damping and Gaussian white noise, resulting in a Markovian dynamics \cite{Starobinsky:1994bd,Biro:1997va,Rischke:1998qy,Bodeker:1998hm,Berera:2004kc,Yokoyama:2004pf,Weenink:2011dd,Nahrgang:2011mg}. This assumes a clear separation of scales between the so-called ``memory time'', usually identified with the short time scale of irrelevant degrees of freedom, and the ``relaxation time'', which characterizes the dynamics of the relevant degrees of freedom \cite{Rau:1995ea,Berera:2007qm}. Existing comparisons between given non-Markovian Langevin dynamics and their Markovian versions, see e.g. \cite{Xu:1999aq,Farias:2009zz,Farias:2009zzz,Fraga:2009} indicate that, indeed, the better the separation of scales, i.e. the weaker the local damping, the better the Markovian approximation.

As mentioned above, memory kernels can in principle be computed in a given theory from, say, loop diagrams. Existing calculations in simple theories with scalar, fermionic and/or gauge fields show that memory kernels and noise correlators typically decay as power law in time, thereby questioning the existence of a memory time and more generally the possibility of a local, Markovian description \cite{Boyanovsky:1995ema,Boyanovsky:1998pg}. Of course, there are situations where one expects a local Markovian Langevin description to be questionable, such as in presence of massless (e.g. Goldstone or gauge) excitations and/or at zero temperature \cite{Boyanovsky:1994me, Boyanovsky:1998pg}. But the absence of a clear memory time scale mentioned here also occur in more standard situations with no infrared issues and at high temperatures, where one would expect a Brownian like motion to be a good description. 

We address this issue in the present paper. We first recall the basic evolution equations for nonequilibrium two-point Green's functions, which involve non-local memory kernels (self-energies), and show their formal equivalence with a fictitious Langevin process. We discuss the basic conditions under which the memory integral and noise kernel can be replaced by local -- mass and damping -- terms and argue that these conditions do not require the existence of a local limit of the kernels themselves. In order to study the validity of a local Markovian limit of the equivalent fictitious Langevin process, we focus on a simple situation where an out-of-equilibrium test field is coupled to a thermal bath with negligible backreaction. The basic QFT equations can then be exactly solved in terms of the equilibrium spectral function of the test field. We find that there exist a characteristic time below which the actual dynamics is indeed Markovian but after which memory effects cannot be neglected, as expected on rather general grounds  \cite{Fonda:1978dk,Peres:1980kp,Nakazato:1995cn}. We investigate this time scale in detail in a simple model with cubic interactions between the test and the bath fields. Finally, we discuss the consistency of the local limit directly in real time, at the level of the equations of motion.

As an illustration, we end this paper with an application of such local description to the physics of decoherence and entropy production in QFT in the so-called incomplete description picture recently advocated in \cite{Campo:2008ju,Giraud:2009tn,Koksma:2009wa,Koksma:2010dt,Koksma:2011dy,Prokopec:2012xv,Koksma:2010zi,Gautier:2011fx,Prokopec:2012xv}. 

\section{The strategy}

\subsection{General setting}

We consider a generic $\mathbb{Z}_2$-symmetric scalar field theory in the symmetric phase.
The nonequilibrium $n$-point functions can be conveniently described by means of time-ordered products of field operators on a closed contour ${\cal C}$ in time \cite{Schwinger:1960qe,Keldysh:1964ud}. The two-point function $G(x,y)=\langle T_\C\varphi(x)\varphi(y)\rangle$ encodes both the statistical  and spectral correlators $F(x,y)=\frac{1}{2}\langle\{\varphi(x),\varphi(y)\}\rangle$ and $\rho(x,y)=i\langle[\varphi(x),\varphi(y)]\rangle$ as
\beq
 G(x,y)=F(x,y)-\frac{i}{2}{\rm sign}_\C(x^0-y^0)\rho(x,y)\,.
\eeq
Here, the brackets $\langle\ldots\rangle$ denote an average with respect to a given density matrix describing the (quantum) state of interest. The common practice is to specify the (out-of-equilibrium) initial conditions at a given finite time $t=0$ \cite{Berges:2003pc,Berges:2004yj}. In the present work, we choose instead to prepare the system in a Gaussian thermal state at $t\to-\infty$ and to kick it away from equilibrium at $t=0$ by means of an appropriate external source. Both procedures are in principle equivalent. The source approach is an efficient device to let the system develops its own non-Gaussian correlations \cite{Serreau:2003wr,Borsanyi:2008ar}. The latter are of particular importance e.g. when it comes to discussing renormalization -- the ultraviolet modes have to be in the correct non-trivial vacuum state \cite{Borsanyi:2008ar,Garny:2009ni}. We shall not be concerned with these issues in this paper and adopting the source approach is only a matter of technical convenience for later calculations. Still we present it in some detail because of its potential interest also for numerical calculations.

We consider a Gaussian nonequilibrium disturbance described by a bilocal source $K$ in the action \footnote{For discussions of non-Gaussian sources, see \cite{Danielewicz:1982kk,Calzetta:1986cq,Berges:2000ur,Garny:2009ni}.}:
\beq
 S_K[\varphi]=S[\varphi]+\frac{1}{2}\int_\C \!d^4\!x\,d^4\!y\,\varphi(x)K(x,y)\varphi(y)\,,
\eeq
where $S$ is the classical action of the system. The Schwinger-Dyson equation for the two-point function on the contour $\C$ reads:  $G^{-1}=G_0^{-1}-\Sigma+K$, where $iG_0^{-1}(x,y)=-(\square+M^2)\delta^{(4)}_\C(x-y)$ is the free propagator for the field of mass $M$ and $\Sigma(x,y)$ its self-energy. Using the standard decomposition \cite{Berges:2003pc,Berges:2004yj}
\bea
 \Sigma(x,y)&=&\Sigma_{0}(x)\delta_\C^{(4)}(x-y)\nn
  &+& \Sigma_F(x,y)-\frac{i}{2}{\rm sign}_\C(x^0-y^0)\Sigma_\rho(x,y)\,.
\eea
and assuming Gaussian conditions at $t\to-\infty$, the Schwinger-Dyson equation translates into the following nonlinear coupled integro-differential equations for the statistical and spectral propagators 
\bea
\left[\square_x +M^2(x)\right]F(x,y)\! &=&\!
- \int_{-\infty}^{x^0}\!\!\!{\rm d}^4z\,\Sigma_{\rho}(x,z)F(z,y) \nonumber \\
&&\!+ \int_{-\infty}^{y^0}\!\!\!{\rm d}^4z\, \Sigma_{F}^K(x,z)
\rho(z,y)\,,
\label{eq:F1}
\\ 
\left[\square_x +M^2(x)\right]\rho(x,y)\! &=&\!
-\int_{y^0}^{x^0}\!\!\! {\rm d}^4z\,
\Sigma_{\rho}(x,z)\rho(z,y)\,,
\label{eq:rho1}
\eea
where $\int_{a}^{b}{\rm d}^4z\equiv\int_{a}^{b}{\rm d}z^0\int \!{\rm d} {\bf z}$, $M^2(x) = M^2 +\Sigma_{0}(x)$ and
\beq
 \Sigma_F^K(x,y)=\Sigma_F(x,y)-K(x,y)\,.
\eeq

For later purposes it proves convenient to rewrite Eqs. \eqn{eq:F1}-\eqn{eq:rho1} by introducing the retarded and advanced propagators $G_R(x,y) = \theta (x^0 - y^0)\rho(x,y)$ and $G_A(x,y) =  G_R(y,x)=-  \, \theta (y^0 - x^0)\rho(x,y)$ and similarly for the self-energies. One gets, from \Eqn{eq:rho1}, 
\beq
G_R ^{-1}(x,y)= \left[\square_x +M^2(x)\right]\delta^{(4)}(x - y)+ \Sigma_R(x,y)
\label{def:GR}
\eeq
and similarly for $G_A(x,y)=G_R(y,x)$. Eqs. \eqn{eq:F1}-\eqn{eq:rho1} can be written \footnote{In general one should add the general solution of the homogeneous equation $G_R*F=0$ but the latter typically decays with time and has an infinite time to decay away in the present set-up where initial conditions are prepared at $t\to-\infty$.}
\bea
F &=& - G_R * (\Sigma_F-K)* G_A \,,
\label{eq:F2} \\
\rho &=& - G_R * \Sigma_\rho* G_A \,,
\label{eq:rho2}
\eea
with $(f*g)(x,y) = \int \!\! {\rm d}^4z \, f(x,z)g(z,y) $. Eqs. \eqn{def:GR}-\eqn{eq:rho2} actually provide an explicit solution if the self-energies are known. In general, however, the latter are non-linear functions of the two-point functions themselves.

The source $K$ serves to prepare out-of-equilibrium conditions near $x^0=y^0=0$.
For instance, we shall consider an instantaneous kick at $x^0=y^0=0$:
\bea
 K(x,y)&=&A({\bf x},{\bf y}) \delta(x^0)\delta(y^0) + B({\bf x},{\bf y})\delta'(x^0)\delta'(y^0) \nn 
 &+&C({\bf x},{\bf y}) \left[\delta'(x^0)\delta(y^0)+\delta(x^0)\delta'(y^0)\right] .
 \label{eq:kick}
\eea
Before the kick, the system has had an infinite amount of time to reach an equilibrium state with parameters (e.g. temperature) determined by the conditions (e.g. energy density) at time $t\to-\infty$:
\beq
 \left.F(x,y)\right|_{x^0,y^0<0}=F_{\rm eq}(x-y)\,.
\eeq
Right after the kick, we write:
\beq
 \left.F(x,y)\right|_{x^0,y^0>0}=F_{\rm neq}(x,y)\,.
\eeq

The functions $A$, $B$ and $C$ in \eqn{eq:kick} are related to initial conditions for the nonequilibrium statistical function $F_{\rm neq}$ \footnote{Initial conditions for the spectral function $\rho$ are fixed by equal-time commutation relations.}. From \Eqn{eq:F2}, we obtain
\beq
\label{eq:fullsol}
 F(x,y)=- \left[G_R*\Sigma_F*G_A\right](x,y)+\theta(x^0)\theta(y^0)F_K(x,y),
\eeq
with
\bea
 F_K(x,y)\!=\!\Big\{\rho(x^0,0)\!\cdot\! A\!\cdot\!\rho(0,y^0)\!+\!\p_u\rho(x^0,u)\!\cdot\! B\!\cdot\!\p_v\rho(v,y^0)&&\nn
 -\p_u\rho(x^0,u)\!\cdot\! C\!\cdot\!\rho(0,y^0)-\rho(x^0,0)\!\cdot\! C\!\cdot\!\p_v\rho(v,y^0)\Big\}_{u=v=0},&&\nn
\eea
where the dot denote a convolution with respect to spatial variables only. Using $\p_{x^0}\rho(x,y)|_{y^0=x^0}=\delta^{(3)}({\bf x}-{\bf y})$ we obtain the relations:
\bea
B({\bf x},{\bf y})&=&  \left. F_K(x,y) \right|_{x^0=y^0=0}, \nn
C({\bf x},{\bf y})&=&\left. \partial_{x^0}  F_K(x,y) \right|_{x^0=y^0=0} ,\\
A({\bf x},{\bf y})&=&\left. \partial_{x^0} \partial_{y^0} F_K(x,y) \right|_{x^0=y^0=0}. \nonumber
\label{CI}
\eea
Somewhat more intuitive expressions of the sources can be obtained by defining
\bea
 F^0_{\rm eq}({\bf x},{\bf y})&=&\left.F(x,y)\right|_{x^0,y^0\to0^-},\nn
 R^0_{\rm eq}({\bf x},{\bf y})&=&\left.\p_{x^0}F(x,y)\right|_{x^0,y^0\to0^-},\\
 K^0_{\rm eq}({\bf x},{\bf y})&=&\left.\p_{x^0}\p_{y^0}F(x,y)\right|_{x^0,y^0\to0^-}\nonumber
\eea
and
\bea
 F^0_{\rm neq}({\bf x},{\bf y})&=&\left.F(x,y)\right|_{x^0,y^0\to0^+},\nn
 R^0_{\rm neq}({\bf x},{\bf y})&=&\left.\p_{x^0}F(x,y)\right|_{x^0,y^0\to0^+},\\
 K^0_{\rm neq}({\bf x},{\bf y})&=&\left.\p_{x^0}\p_{y^0}F(x,y)\right|_{x^0,y^0\to0^+}.\nonumber
\eea
Assuming that the function $G_R*\Sigma_F*G_A$ is continuous at $x^0=y^0=0$, one has 
\bea
 B({\bf x},{\bf y})&=&F^0_{\rm neq}({\bf x},{\bf y})-F^0_{\rm eq}({\bf x},{\bf y}),\nn
 C({\bf x},{\bf y})&=&R^0_{\rm neq}({\bf x},{\bf y})-R^0_{\rm eq}({\bf x},{\bf y}),\\
 A({\bf x},{\bf y})&=&K^0_{\rm neq}({\bf x},{\bf y})-K^0_{\rm eq}({\bf x},{\bf y}).\nonumber
\eea

\subsection{Equivalent Langevin process}

The evolution equations \eqn{def:GR}-\eqn{eq:rho2} are formally equivalent to a fictitious Langevin process $\varphi^\xi$ driven by a random noise $\xi$ \footnote{This (exact) equivalence has been repeatedly noticed in the literature for the case where the field of interest is linearly coupled to an environment on which it does not backreact, see e.g.  \cite{Feynman:1963fq, Boyanovsky:2004dj, Drewes:2012qw}. In more general cases, the influence functional approach of Feynman and Vernon allows one to derive approximate Langevin equations \cite{Feynman:1963fq, Rischke:1998qy,Berera:2007qm}. To our knowledge, the -- formal but exact -- equivalence described in the present paper has not been pointed out before.}:
\beq
 \left[\square_x +M^2(x)\right]\varphi^\xi(x)+\int {\rm d}^4 z\, \Sigma_R(x,z)\varphi^\xi(z) = \xi(x) + \Xi(x),
\label{eq:noise_osc}
\eeq  
where $\Xi$ is a fluctuating disturbance source which serves to send the system away from equilibrium at a given finite time, in the same spirit as the source $K$ above. The random variables $\xi$ and $\Xi$ are uncorrelated. 

Using \eqn{def:GR}, \Eqn{eq:noise_osc} can be rewritten as $G_R ^{-1}*\varphi^\xi = \xi + \Xi$, whose solution is formally given by: 
\beq
\varphi^\xi(x) = \varphi_0(x)+\int  {\rm d}^4 z\,G_R(x,z)\left[\xi(z) + \Xi(z)\right]\,,
\label{eq:sol}
\eeq 
where $\varphi_0$ is the general solution of the homogeneous equation $G_R^{-1}*\varphi_0=0$.

We shall denote the average over realizations of the noise $\xi$ with an overline and the average over the source $\Xi$ with double brackets. The original quantum average is identified with $\langle\ldots\rangle\equiv\overline{\langle\langle\ldots\rangle\rangle}$. Setting $\langle\langle\Xi(x)\rangle\rangle=0$ and $\overline{\xi(x)}=0$ and choosing initial conditions such that $\varphi_0(x)=0$ guarantees that $\langle\varphi(x)\rangle\equiv\overline{\langle\langle\varphi^\xi(x)\rangle\rangle}=0$. To obtain the evolution equation for the two-point function
\beq
 \label{def:Fnoise}
F(x,y) \equiv \overline{\langle\langle\varphi^\xi(x)\varphi^{\xi}(y)\rangle\rangle}\,,
\eeq
we write \eqn{eq:noise_osc} as
\beq
\label{eq:noise_bis}
 \int {\rm d}^4 z\, G_R^{-1}(x,z)\varphi^\xi(z)\varphi^{\xi}(y) = \left[\xi(x)+\Xi(x)\right]\varphi^\xi(y)\,.
\eeq
Inserting \Eqn{eq:sol} on the right-hand-side and identifying 
\beq
 \Sigma_F(x,y)=-\overline{\xi(x)\xi(y)}\quad{\rm and}\quad K(x,y)=\langle\langle\Xi(x)\Xi(y)\rangle\rangle\,,
\eeq 
one recovers \Eqn{eq:F2}. 
 
The equivalence is complete provided the sources are chosen in an appropriate way. For the case of an instantaneous kick,
\beq
\Xi(x) = \Upsilon({\bf x})\delta(x^0) + \Pi({\bf x})\delta^\prime(x^0)\,,
\eeq
one gets
 \bea
B({\bf x},{\bf y})&=& \langle\langle\Pi({\bf x})\Pi({\bf y})\rangle\rangle\,,\nn
C({\bf x},{\bf y})&=& \frac{1}{2} \langle\langle\Upsilon({\bf x})\Pi({\bf y})+\Pi({\bf x})\Upsilon({\bf y})\rangle\rangle\,, \\ 
A({\bf x},{\bf y})&=&  \langle\langle\Upsilon({\bf x})\Upsilon({\bf y}) \rangle\rangle\,,  \nonumber
\label{CI_bis}
\eea
with $ \langle\langle\Upsilon({\bf x})\Pi({\bf y})-\Pi({\bf x})\Upsilon({\bf y})\rangle\rangle=0$.

It is worth emphasizing that, in general, the noise correlator $\Sigma_F$ and the memory kernels $\Sigma_\rho$ depend nonlinearly on the correlators $F$ and $G_R$ and are thus nonlinear functions of the field $\varphi^\xi$ and the noise $\xi$ and their derivatives. In particular, this encodes all sorts of additive and/or multiplicative noises as well as linear and/or nonlinear damping terms that one encounters in usual (mostly perturbative) derivations of Langevin equations e.g. based on the influence functional technique \cite{Berera:2004kc,Yokoyama:2004pf}. 
 
\subsection{Approximation strategy}
\label{sec:as}

We now examine sufficient conditions under which the memory integral in the exact nonlocal Langevin equation \eqn{eq:noise_osc} can be approximately described by a local damping term. We shall see that there is no need a priori for assuming a finite memory time. For simplicity, we shall specialize to spatially homogeneous and isotropic situations. In particular the source $K$ and, consequently, all other two-point functions in the problem can be spatially Fourier transformed:
\beq
 K(x,y)=\int\frac{d^3p}{(2\pi)^3}e^{i{\bf p}\cdot({\bf x}-{\bf y})}K_p(x^0,y^0)
\eeq
and similarly for all other two-point functions. The relevant equations read
\bea
&&\left[\p_t^2 + \omega_p^2(t)\right]F_p(t,t^\prime) + \int_{-\infty}^t \!\!{\rm d}u \,\Sigma_p^{\rho}(t,u) F_p(u,t')  \nn 
&&\hspace{2.2cm}= -\int_{-\infty}^{+\infty} \!\!\!{\rm d}u \, \Sigma_p^{F,K}(t,u)G^A_p(u,t')
\label{eq:Ffourier}
\eea
and
\beq
\left[\p_t^2 + \omega_p^2(t)\right]G^{R}_p(t,t^\prime)+ \int_{-\infty}^t \!\!\!{\rm d}u\, \Sigma_p^{\rho}(t,u) G^{R}_p(u,t') = 
\delta(t-t'),
\label{eq:GRAfourier}
\eeq
where $\omega_p(t)=\sqrt{p^2+M^2(t)}$ and $\Sigma_p^{F,K}=\Sigma_p^{F}-K_p$.

In generic theories, the memory kernel $\Sigma_p^\rho(t,t')$ is typically a decreasing function of $t-t'$. For instance, writing the interaction term of the field degree of freedom of interest $\varphi_p(t)$ as $S_{\rm int}=\int_{t,p}\varphi_p(t)J_p(t)$, where the current $J_p$ involves other degrees of freedom, possibly including $\varphi_{p'\neq p}$, one has, at lowest order in the interaction,
\beq
 \Sigma_p^\rho(t,t')\propto\langle[J_p(t),J_p(t')]\rangle\,.
\eeq
Such current-current correlator typically exhibits rapid -- e.g. power law \cite{Moore:2002mt,Boyanovsky:1995ema,Boyanovsky:1998pg} -- decay at large $t-t'$ and the memory integrals in Eqs. \eqn{eq:Ffourier} and \eqn{eq:GRAfourier} are dominated by their upper bound.

As for the functions $F_p$ and $G_p^R$, they are generically characterized by various -- possibly time-dependent -- time scales: an oscillation frequency $\epsilon_p(t)$, the damping scale in the relative time $t-t'$ and the typical scale of the nonequilibrium evolution of equal-time correlators -- the last two might actually be power laws. The most common situation -- e.g. for weakly coupled (effective) theories --  is that the former is the shortest time scale in the problem. We shall assume a clear separation of scales, where the oscillation frequency is always much shorter than the damping or nonequilibrium scales.

Now, if the range of integration which dominates the memory integrals is short compared to the above-mentioned damping and nonequilibrium scales, one can neglect the latter under these integrals and replace the full functions $F_p$ and $G_p^R$ by their short-time oscillating part around the upper bound of the integral, e.g. \footnote{This is somewhat similar to so-called harmonic, or one-frequency ansatz employed in Refs. \cite{Greiner:1996dx,Greiner:1998vd,Rischke:1998qy}. Here, the convolution with the memory kernel $\Sigma_\rho$ controls the convergence of the integral and gives better control of the approximation.}:
\bea
\label{eq:harmonic}
 F_p(u,t') &\approx& F_p(t,t') \cos \epsilon_p(t)(u-t)\nn
 &-&\frac{\partial_tF_p(t,t')}{\epsilon_p(t)}\sin \epsilon_p(t)(u-t)\,,
\eea
with some relative error controlled by the ratios of either the damping or nonequilibrium scales over the oscillation frequency. The memory integral in \eqn{eq:Ffourier} can thus be approximated as
\beq
\int_{-\infty}^t \!\!\!\!{\rm d}u \,\Sigma_p^{\rho}(t,u) F_p(u,t')\approx \delta\epsilon_p^2(t) F_p(t,t^\prime)\!+\!2\gamma_p(t)\partial_tF_p(t,t'),
 \label{eq:approxlocal}
\eeq
with
\beq
 \delta\epsilon_p^2(t)\equiv {\rm Re}\tilde\Sigma_p^R(t;\epsilon_p(t))
 \label{def:delta}
\eeq
and
\beq
 \gamma_p(t)\equiv -\frac{{\rm Im}\tilde\Sigma_p^R(t;\epsilon_p(t))}{2\epsilon_p(t)}\,,
  \label{def:gamma}
\eeq
where we used the definition $\Sigma_p^R(t,t')=\theta(t-t')\Sigma_p^{\rho}(t,t')$ and introduced the mixed time-frequency representation
\beq
 \Sigma_p^R(t,t')=\int \frac{d\omega}{2\pi} \, e^{-i\omega (t-t')}\tilde\Sigma_p^R(t;\omega)\,.
\eeq
Similar manipulations hold for the memory integral in \Eqn{eq:GRAfourier}.

If the above conditions are met, the equation for $F_p$ and $G_p^R$ thus read
 \bea
&&\left[\p_t^2 +  2 \gamma_p(t) \p_t + \epsilon_p^2(t)\right]F_p(t,t^\prime) = \nn
&&\hspace{2.3cm} -\int_{-\infty}^{+\infty}\!\!\!{\rm d}u \, \Sigma_p^{F,K}(t,u)G_p^A(u,t')
\label{eq:Langevin}
\eea
and
 \beq
 \left[\p_t^2 +  2 \gamma_p(t) \p_t +\epsilon_p^2(t) \right]G_p^{R,A}(t,t^\prime) =\delta(t-t')\,,
\eeq
where we defined $\epsilon_p^2(t)=\omega_p^2(t) + \delta\epsilon_p^2(t)$. 
Using similar manipulations as in the previous subsection, it is easy to check that these equations also have an equivalent Langevin process which reads:
\beq
\label{eq:locallangevin}
 \left[\p_t^2 +  2 \gamma_p(t) \p_t +\epsilon_p^2(t) \right]\varphi^\xi_p(t)=\xi_p(t)+\Xi_p(t)\,.
\eeq
Note that at this stage, the memory kernel has been replaced by a local damping term but the noise correlator can still be non-trivial, i.e. non-Markovian. Note also that, here, the damping rate and the noise correlator are still, in general, complicated non-linear functions of the propagators.

\section{Exactly solvable example: near stationary system}
\label{sec:exact}

We now want to investigate in more details the condition under which the above local approximation is valid. In this section we specialize to a simple situation where the out-of-equilibrium degrees of freedom weakly interact with a stationary -- e.g. thermal -- background with negligible backreaction. In this case, the full nonequilibrium dynamics can be exactly solved \cite{Boyanovsky:2004dj,Anisimov:2008dz,Garbrecht:2011xw} and one can check whether the exact solution can be described by a Langevin dynamics with local damping.

 \subsection{Exact solution}
 
The assumption of negligible backreaction means that the self-energies $\Sigma_{F,\rho}$ are determined by the stationary background and are thus time-translation invariant: $\Sigma_p^{F,\rho}(t,t')\equiv\Sigma_p^{F,\rho}(t-t')$. It follows that the spectral function is also time-translation invariant: $\rho_p(t,t')=\rho_p(t-t')$ \cite{Anisimov:2008dz}. Indeed, using the time-translation invariance of $\Sigma_p^\rho$, one easily checks that the equation of motion for $\rho_p(t,t')$, see e.g. \eqn{eq:GRAfourier}, only explicitly involves the time difference $t-t'$ and thus possesses time-translation invariant solutions. Moreover, since the equal-time commutation relations, which determine the initial conditions for $\rho_p$, are preserved under time-evolution, the only possible solution has this symmetry. The equation of motion for $\rho_p(t-t')$ reads:
\beq
\label{eq:rhop}
\left[\partial_t^2+\omega_p^2\right]\rho_p(t)+\int_{0}^{t}\!\! d\tau \,\Sigma_p^{\rho}(t-\tau)\rho_p(\tau)=0\,,
\eeq
with $\rho(0)=0$ and $\dot\rho(0)=1$. It follows that the functions $G_p^{R,A}$ and, in turn, $F_{\rm eq}=-G_R*\Sigma_F*G_A$ are also time-translation invariant. 

Introducing the frequency representation 
\beq
 \rho_p(t)=\int\frac{d\omega}{2\pi}\,e^{-i\omega t}\tilde\rho_p(\omega)
\eeq
and similarly for all other time-translation invariant functions $G_p^{R,A}$ and $\Sigma_p^{F,\rho,R,A}$, one readily obtain
\beq
 \tilde G_p^R(\omega)=\frac{1}{-(\omega+i0^+)^2+\omega_p^2+\tilde \Sigma_p^R(\omega+i0^+)}\,,
 \label{eq:retarded}
\eeq
as well as
\bea
\tilde F_p^{\rm eq}(\omega) &=& - \big| \tilde G_p^R(\omega) \big|^2 \tilde \Sigma_p^{F}(\omega) \,,
\label{eq:Ff} \\
\tilde\rho_p(\omega) &=& - \big| \tilde G_p^R(\omega) \big|^2 \tilde \Sigma_p^\rho (\omega) \,.
\label{eq:rhof}
\eea
It is useful to recall that $\tilde\rho_p(\omega)=2i\,{\rm Im}\tilde G_p^R(\omega)$ and similarly for the self-energy components \footnote{It is also useful to recall that $\tilde G_p^R(-\omega)=[\tilde G_p^A(\omega)]^*$, $\tilde\rho_p(-\omega)=-\tilde\rho_p(\omega)=[\tilde\rho_p(\omega)]^*$, $\tilde F^{\rm eq}_p(-\omega)=\tilde F^{\rm eq}_p(\omega)=[\tilde F^{\rm eq}_p(\omega)]^*$ and similarly for the various self-energy components.}. Notice the detailed balance relation
\beq
 \tilde F^{\rm eq}_p(\omega)\tilde\Sigma_p^\rho(\omega)=\tilde\rho_p(\omega)\tilde\Sigma_p^F(\omega)\,,
\eeq
characteristic of stationary systems. In the following we assume a thermal background to fix the ideas although the argument does not depend on this assumption. In thermal equilibrium at temperature $T=1/\beta$, the fluctuation-dissipation relation
\beq
\label{eq:FDrel}
 \tilde F^{\rm eq}_p(\omega)=-i\Big(n(\beta\omega)+{1\over2}\Big)\tilde\rho_p(\omega)
\eeq
implies
\beq
\label{eq:FDrel2}
 \tilde\Sigma_p^F(\omega)=-i\left(n(\beta\omega)+{1\over2}\right)\tilde\Sigma^\rho_p(\omega)\,,
\eeq
with $n(x)=1/(\exp(x)-1)$ and thus there is only one independent function, e.g. $\tilde\rho_p(\omega)$.

For an instantaneous kick at $t=t'=0$ as in \eqn{eq:kick}:
\bea
  K_p(t,t')&=&A_p\delta(t)\delta(t')+ B_p\delta'(t)\delta'(t')\nn 
 &+&C_p\left[\delta'(t)\delta(t')+\delta(t)\delta'(t')\right]\,,
\eea
the complete nonequilibrium solution for $F_p$ reads, see Eqs. \eqn{eq:fullsol}-\eqn{CI},
\beq
\label{eq:fullsolfourier}
 F_p(t,t') = F^{\rm eq}_p(t-t')+\theta(t)\theta(t')F^{K}_p(t,t')\,,
\eeq
with
\bea
 F^{K}_p(t,t')&=&A_p\rho_p(t)\rho_p(t')  + B_p\dot\rho_p(t)\dot\rho_p(t')\nn
 \label{eq:relaxation}
 &+ &C_p\left[\dot\rho_p(t)\rho_p(t')+\rho_p(t)\dot\rho_p(t')\right]\,,
\eea
where
\bea
B_p &=&  \left. F_p^{K}(t,t') \right|_{t=t'=0},\nn
\label{eq:CImom}
C_p &=& \left. \partial_{t}  F_p^{K} (t,t') \right|_{t=t'=0}, \\
A_p &=&  \left.\partial_{t} \partial_{t'}  F_p^{K}(t,t')  \right|_{t=t'=0}.\nonumber 
\eea

Thus we see that, in the simple situation considered here, the nonequilibrium dynamics is fully described by the second term on the right-hand-side of \Eqn{eq:fullsolfourier}: The momentum modes of interest get kicked away from equilibrium at $t=0$ and the relaxation toward equilibrium, described by \Eqn{eq:relaxation}, is completely encoded in the (equilibrium) spectral function $\rho_p(t)$. 

 \subsection{Breit-Wigner approximation}
 
In general, the spectral function $\tilde\rho_p(\omega)$ may exhibits single-particle poles as well as branch cuts in the complex frequency plane, corresponding to multiparticle processes or Landau damping, see e.g. \cite{Boyanovsky:1995ema,Boyanovsky:1998pg,Drewes:2010pf}. For instance, near the poles, it takes the usual Breit-Wigner form \footnote{There is a similar expression near the symmetric pole $\omega\to-\epsilon_p$ -- recall that $\tilde\rho_p(-\omega)=-\tilde\rho_p(\omega)$.}
\beq
\label{eq:BW}
 \tilde\rho_p(\omega\to\epsilon_p)\approx\frac{Z_p}{\epsilon_p}\,\frac{i\gamma_p}{(\omega-\epsilon_p)^2+\gamma_p^2}\,,
\eeq
where we assume $\gamma_p/\epsilon_p\ll1$. Here,
\beq
\label{eq:fieldrenorm}
  Z_p^{-1}=1-\frac{d}{d\omega^2}{\rm Re}\tilde\Sigma^R_p(\omega)\Big|_{\omega=\epsilon_p}
 \eeq
and the quasiparticle energy $\epsilon_p$ and width $\gamma_p$ are determined from
\beq
 \epsilon_p^2=\omega_p^2+{\rm Re}\,\tilde\Sigma^R_p(\epsilon_p) 
\eeq
and
\beq
 \gamma_p=-Z_p\frac{{\rm Im}\tilde\Sigma^R_p(\epsilon_p)}{2\epsilon_p}\,.
 \label{def:gammabis}
\eeq

Poles lead to exponential damping of the Fourier transform $\rho_p(t)$ at large time, whereas branch cuts result in power law behavior \cite{Boyanovsky:1995ema,Boyanovsky:1998pg}. The large-time behavior is, typically,
\beq
\label{eq:rhopow}
 \rho_p(t)\approx Z_p\frac{\sin\epsilon_pt}{\epsilon_p}e^{-\gamma_p|t|}+\rho_p^{\rm pow}(t)\,,
\eeq
where the last term denotes the power law contribution. Using \eqn{eq:FDrel} one can obtain the corresponding large-time behavior of the equilibrium statistical function $F_p^{\rm eq}(t-t')$:
\beq
\label{eq:Fpow}
F_p^{\rm eq}(t) \approx \kappa_p\left(\cos\epsilon_pt+\frac{\gamma_p}{\epsilon_p}\sin\epsilon_p|t|\right) e^{- \gamma_p| t|} + F_p^{\rm pow}(t)\,,
\eeq
with
\beq
 \kappa_p=\frac{Z_p}{\epsilon_p}\left(n\left(\beta\epsilon_p\right)+{1\over2}\right).
\eeq

The power law contributions to both $\rho_p(t)$ and $F_p^{\rm eq}(t)$ are due to features of the self-energies $\tilde\Sigma_p^{F,\rho}(\omega)$ and their amplitudes are thus governed by the strength of the interaction with the thermal bath, i.e. by some power of the relevant coupling constant. This is to be contrasted with the pole contributions, where the interaction strength essentially controls the decay rate $\gamma_p$ but only gives corrections to the amplitude. Thus, for weak coupling, there exist a range of time during which the exponentially decaying contributions dominate over the power laws. Only at very late time is the dynamics described by the latter. 

Writing schematically the power law contributions as $\sim {\cal A}/(\mu t)^{\nu}$, with ${\cal A}$ an amplitude proportional to some positive power of the relevant coupling constant or, equivalently, of the damping rate $\gamma_p$, $\mu$ some scale of the problem at hand and $\nu$ a given exponent \footnote{All these quantities may in principle depend on the momentum $p$. We omit the $p$ index for simplicity.}, one finds that the power law contributions are of relative order $\gamma_p/\epsilon_p$ for times 
\beq 
\label{eq:trange}
 \gamma_pt\lesssim\ln \left\{\left({\mu\over\gamma_p}\right)^{\!\nu}\!{\gamma_p\over\epsilon_p^2{\cal A}}\right\}\,.
 \eeq 
 In this time range, the nonequilibrium relaxation \eqn{eq:relaxation} is also essentially exponential:
\beq
\label{eq:exprel}
 F_{p}^{K}(t,t')\approx F^0_p(t,t')e^{-\gamma_p(t+t')}\,,
\eeq
with
\beq
 F^0_p(t,t')\!= \!a^+_p\!\cos\epsilon_p(t-t')+ a^-_p\!\cos\epsilon_p(t+t')+ a^0_p\sin\epsilon_p(t+t'),
\eeq
where the constant $a^\pm_p$ and $a^0_p$ are related to initial conditions \eqn{eq:CImom}
\bea
 a^\pm_p &=&  \frac{Z_p^2}{2}\left(B_p\pm \frac{A_p-2\gamma_pC_p+\gamma_p^2 B_p}{\epsilon_p^2}\right) \,,\nn
 a^0_p &=&  Z_p^2\frac{C_p-\gamma_pB_p}{\epsilon_p} \,.
\label{CI_again}
\eea

Thus the complete nonequilibrium solution takes a remarkably simple form in the time range \eqn{eq:trange} \footnote{The solution obtained here generalizes that of Ref. \cite{Garbrecht:2011xw}, which corresponds to the particular nonequilibrium conditions $a_p^-=a_p^0=0$.}.  In the weak coupling limit both $\gamma_p/\mu_p$ and $\epsilon_p{\cal A}_p$ are small and the latter can be relatively long. In that case, the deviation from equilibrium \eqn{eq:exprel} as well as the equilibrium unequal-time correlators \eqn{eq:rhopow} and \eqn{eq:Fpow}  have essentially completely decayed when the power law contributions start becoming important and most of the relevant dynamics is thus well described by the exponential -- pole -- contributions. As we now discuss, the latter can be described by an equivalent Markovian Langevin dynamics.

\subsection{Markovian Langevin dynamics}
\label{sec:MLd}

In the time range  \eqn{eq:trange}, the dynamics is dominated by the poles of the retarded Green's function \eqn{eq:retarded}. Those which are closest to the real axis give the dominant late-time contribution. Assuming that the dynamics is driven by these poles and that the Breit-Wigner approximation \eqn{eq:BW} is justified, one can make the replacement $\tilde G_p^R(\omega)\to\tilde G_{{\rm eff},p}^R(\omega)$ with \footnote{We choose this ansatz for convenience although the poles are slightly displaced by a negligible amount $\sim\gamma_p^2/\epsilon_p$ along the real axis. In particular, it exactly  reproduces \Eqn{eq:BW} -- with no $\gamma_p^2$ correction -- when $\omega\to\epsilon_p$.}
\beq
\label{eq:GReff}
 \tilde G_{{\rm eff},p}^R(\omega)=\frac{Z_p}{-(\omega+i\gamma_p)^2+\Omega_p^2},
\eeq
where $\Omega_p^2=\epsilon_p^2-\gamma_p^2\approx\epsilon_p^2$. The corresponding spectral function reads
\beq
\label{eq:rhoeff}
 \tilde\rho_{{\rm eff},p}(\omega)=\frac{4iZ_p\gamma_p\omega}{(\omega^2-\epsilon_p^2)^2+4\gamma_p^2\omega^2}\,,
\eeq
which reprocuces \Eqn{eq:BW} near the poles. The Fourier transform reads
\beq
 \rho_{{\rm eff},p}(t)=Z_p\frac{\sin\epsilon_pt}{\epsilon_p}e^{-\gamma_p|t|}\,.
 \label{sol:rhomarkov}
\eeq

The function \eqn{eq:GReff} is the frequency space representation of the retarded Green's function of the operator $Z_p^{-1}\!\left[\p_t^2 +  2 \gamma_p \p_t +\epsilon_p^2\right]\delta(t-t')$. It follows that the nonequilibrium relaxation \eqn{eq:exprel} can be described by a Langevin process with local damping:
\beq
\label{eq:langevineff}
 \left[\p_t^2 +  2 \gamma_p \p_t +\epsilon_p^2 \right]\varphi^\xi_p(t)=\xi_p(t)+\Xi_p(t)
\eeq
with appropriate sources $\Xi_p(t)$. The equilibrium state is characterized by the $\xi_p(t)$ correlator. A Gaussian white noise 
\beq
\label{eq:noiseeff}
 \overline{\xi_p(t)\xi_p(t')}=Z_p^{-1}\alpha_p\delta(t-t')
\eeq
gives, after simple calculations,
\beq
 \tilde F_{\rm eff}^{\rm eq}(\omega)=\frac{Z_p\alpha_p}{(\omega^2-\epsilon_p^2)^2+4\gamma_p^2\omega^2}
\eeq
or, equivalently,
\beq
\label{eq:Feqeff}
 F^{\rm eq}_{{\rm eff},p}(t)=\frac{Z_p\alpha_p}{4\gamma_p\epsilon_p^2}\left(\cos\epsilon_pt+\frac{\gamma_p}{\epsilon_p}\sin\epsilon_p|t|\right)e^{-\gamma_p|t|}\,.
\eeq
Choosing
\beq
\label{eq:FDnoisedamping}
 \alpha_p = 4\gamma_p \epsilon_p \left(n(\beta\epsilon_p) +{1\over2}\right)\,,
\eeq
one reproduces the exponential term in \eqn{eq:Fpow}. Thus the simple Langevin dynamics \eqn{eq:langevineff}-\eqn{eq:noiseeff} with local damping and white (Markovian) noise related by \eqn{eq:FDnoisedamping} correctly describes the full nonequilibrium dynamics of the previous subsection in the time range \eqn{eq:trange}.

In the present case, both the local damping and the Markovian nature of the noise follow from the Breit-Wigner approximation and the assumption of a thermal bath. Indeed the Breit-Wigner ansatz \eqn{eq:GReff}, see also \eqn{eq:BW}, which leads to local damping $\sim\gamma_p\partial_t$, implicitly assumes that $\tilde\Sigma_p^\rho(\omega)$ is sufficiently smooth on a range $|\epsilon_p-\omega|\lesssim\gamma_p$. For the case of a thermal bath, the fluctuation-dissipation relation \eqn{eq:FDrel2} guarantees that $\tilde\Sigma_p^F(\omega)$ has the same property. Assuming that the poles give the dominant contribution to $F_p^{\rm eq}(t)$ amounts to neglecting the frequency dependence of $\tilde\Sigma_p^F$ around the pole and replacing $\tilde\Sigma_p^F(\omega)\approx\tilde\Sigma_p^F(\epsilon_p)$ in \Eqn{eq:Ff}. In terms of the equivalent Langevin description, this corresponds to white noise. We see that, in principle, one may have local damping but non-local, colored noise e.g. in the case of some non-thermal background.

Finally, let us recall that the relation \eqn{eq:FDnoisedamping} follows from the on-shell fluctuation-dissipation relation \eqn{eq:FDrel} \cite{Rischke:1998qy}. Indeed, the effective Langevin dynamics corresponds to the following effective self-energy kernels:
\beq
  \tilde\Sigma_{{\rm eff},p}^R(\omega)=\delta\epsilon_p^2-2iZ_p^{-1}\gamma_p\omega\,,
\eeq
with $\delta\epsilon_p^2={\rm Re}\tilde\Sigma_p^R(\epsilon_p)$ and
\beq
 \tilde\Sigma_{{\rm eff},p}^F(\omega)=-Z_p^{-1}\alpha_p=\tilde\Sigma_{p}^F(\epsilon_p)\,.
\eeq
In particular, one has $\tilde\Sigma_{{\rm eff},p}^\rho(\omega)=-4iZ_p^{-1}\gamma_p\omega$
and thus, on-shell,
\beq
 \tilde\Sigma_{{\rm eff},p}^\rho(\epsilon_p)=-4iZ_p^{-1}\gamma_p\epsilon_p=\tilde\Sigma_{p}^\rho(\epsilon_p)\,.
\eeq
\Eqn{eq:FDnoisedamping} then directly follows from \eqn{eq:FDrel2} taken at $\omega=\epsilon_p$. It generalizes the standard relation $\alpha_p=4\gamma_pT$ valid at high temperature $\beta\epsilon_p\ll1$ \cite{Greiner:1996dx,Greiner:1998vd,Rischke:1998qy}. Equivalently, the equilibrium correlators \eqn{eq:rhoeff} and \eqn{eq:Feqeff} also satisfy the fluctuation-dissipation relation on-shell. Here, the fluctuation-dissipation relations \eqn{eq:FDrel}-\eqn{eq:FDrel2} only hold on-shell, as a consequence of the assumption that the dynamics is governed by the poles. 

To close this subsection, let us remark that the effective memory kernel and noise correlator are completely local:
 \beq
 	\Sigma_{{\rm eff},p}^F(t) = - Z_p^{-1} \alpha_p \delta(t)
\eeq
and
\beq
 \Sigma_{{\rm eff},p}^R(t)=\delta\epsilon_p^2\delta(t)+2Z_p^{-1} \gamma_p\delta'(t)\,.
\eeq
As explained in the introduction, this seems in contradiction with actual calculations of such functions in various models, which typically give power laws in time \cite{Boyanovsky:1995ema,Boyanovsky:1998pg}. We present an explicit example in the next section. Then, we study in Sec. \ref{sec:check} how the apparent contradiction is resolved.

\section{An explicit example}
\label{sec:explicit}

To illustrate the point, we consider a simple model with non-trivial dissipation \cite{Boyanovsky:1995ema,Boyanovsky:1998pg,Anisimov:2008dz} where the (thermal) background is represented by a scalar field $\chi$ of mass $m$ interacting with the system field $\varphi$ via a $\varphi\chi^2$ interaction:    
\bea
S &= & -\int \! {\rm d}^{4}x \left\{  {1\over2}\varphi\left(\square+M^2\right)\varphi+ {g\over2} \varphi\chi^2\right.\nn
&&\hspace{1.3cm}\left.+{1\over2}\chi\left(\square+m^2\right)\chi+ V(\chi)\right\} .
\eea
We assume weak enough interaction between the two fields and strong enough $\chi$ self interactions such that back reaction can be neglected and $\chi$ remains in thermal equilibrium at temperature $T$. We shall compute the relevant self-energies in perturbation theory at lowest non-trivial order, assuming a free thermal gas for the $\chi$ field. The spectral and statistical components of the space-time translation-invariant self-energy of the system field $\Sigma(x,y)\equiv\Sigma(x-y)$ read, at one-loop,
\bea
	\label{eq:sigmarho}
	\Sigma_{\rho}(x)    &=& -g^2 F_\chi (x)\rho_\chi(x)\,,\\
 	\Sigma_{F}(x)   &=& -\frac{g^2}{2} \left[F_\chi^2(x) - \frac{1}{4}\rho_\chi^2(x)\right] \,,
	\label{eq:sigmaf}
\eea
where $F_\chi$ and $\rho_\chi$ denote the free statistical and spectral two-point function of the field $\chi$, respectively. Their spatial Fourier transforms read ($\omega_p = \sqrt{p^2 + m^2} $)
\beq
	F_p^\chi(t)    = \left(n(\beta\omega_p) + \frac{1}{2}\right) \frac{\cos\omega_p t}{\omega_p} \,,\quad
	\rho_p^\chi(t)    =  \frac{\sin\omega_pt}{\omega_p}   \,.
	\label{freefield}
\eeq
For simplicity, we consider only the self-energies \eqn{eq:sigmarho}-\eqn{eq:sigmaf} at zero momentum in this section. The discussion is similar for non-vanishing momentum but is a bit more involved due to Landau damping effects which contribute extra features in the complex frequency plane. 

Before going on, we mention a peculiarity of the present model with cubic interaction vertex: the zero mode of the statistical self-energy $\Sigma_{p=0}^{F}(t)$ acquires a time-independent contribution and thus does not decay to zero at large times. This is most easily seen directly in the mixed time-momentum representation:
\beq
\label{eq:sigmapt}
 \Sigma_{p=0}^{F}(t)=-\frac{g^2}{2} \int\frac{d^3q}{(2\pi)^3}\left\{\left[F_q^\chi(t)\right]^2 - \frac{1}{4}\left[\rho_q^\chi(t)\right]^2\right\}.
\eeq
The constant contribution arises from the first term on the right-hand-side, using \eqn{freefield} and $2\cos^2\omega_qt=\cos2\omega_qt+1$ \footnote{We emphasize that this constant only appears for the mode $p=0$. Also, it is a particular feature of the present model with cubic interaction vertex. For instance, it is easy to check that there is no such constant term for the lowest non-trivial contribution to damping (a two-loop setting-sun diagram) for a theory with quartic interaction vertex.}. The actual value of this constant is not important for the present discussion and we refer the interested reader to Appendix \ref{appsec:kappa} for more details. In the following we denote it by $\Sigma_{p=0}^{F}(t\to\infty)=\sigma_\infty$ (non constant contributions vanish at large time, see below). In the rest of this section, we omit the index $p=0$ for simplicity.

The self-energy \eqn{eq:sigmarho}-\eqn{eq:sigmaf} read, in momentum-frequency space \cite{Boyanovsky:2004dj,Drewes:2010pf}
\beq
	\label{freefieldsigmarho} 
	\tilde\Sigma_{\rho}(\omega) =  -\frac{i g^2}{4\pi} \theta (\omega^2 - 4m^2)\sqrt{1 - \frac{4m^2}{\omega^2}} \left[n\left({\beta\omega\over2}\right) + {1\over2}\right]
\eeq
and (see Appendix \ref{appsec:kappa} for a discussion of the $\delta(\omega)$ term)
\beq
	\label{freefieldsigmaF}
	\tilde\Sigma_{F}(\omega)    =  -i\left(n(\beta\omega) + {1\over2}\right)\tilde\Sigma_{\rho}(\omega)+\sigma_\infty\delta(\omega)\,.
\eeq
In \Eqn{freefieldsigmarho}, one recognizes the two particle threshold at $\omega\ge2m$ which implies that  the on-shell damping rate \eqn{def:gammabis} is non-zero only if $M > 2m$. The square root factor is the standard threshold function for a two body decay and is responsible for the branch cut singularity in the complex frequency plane, which governs the late time power law behavior. Finally the factor  $n(\beta\omega)+1/2$ is due to Bose enhancement and yields poles at imaginary frequencies $\beta\omega_n=4ni\pi$, with $n\in\mathbb{Z}$. 

The late time behavior of the Fourier transforms $\Sigma_{F,\rho}(t)$ can be obtained by standard contour integration techniques. For $m\neq0$, the regime $Tt \gg 1 $ is dominated by the two particle threshold at $\omega=2m$. We obtain,
see also \cite{Boyanovsky:1995ema,Boyanovsky:1998pg,Boyanovsky:2004dj},
\bea
		\Sigma_{\rho}(t) & \approx& \sigma^{\rm pow}_\rho(t)=- a_\rho\frac{\cos{(2mt + \pi/4)}}{(mt)^{3/2}},\\
		\Sigma_{F}(t) & \approx&  \sigma^{\rm pow}_F(t)+\sigma_\infty=a_F \frac{\sin{(2mt + \pi/4)}}{(mt)^{3/2}}+\sigma_\infty,\nn
		\label{latetime}
\eea
with
\bea
 a_\rho&=& \frac{g^2m}{8\pi^{3/2}}\left(n\left({\beta m}\right)+ {1\over2}\right),\\
 a_F&=&\left(n(2\beta m)+ {1\over2}\right)a_\rho\nn
 \label{eq:ttt}
 &=& \frac{g^2m}{16\pi^{3/2}}\left[\left(n\left({\beta m}\right)+{1\over2}\right)^2+{1\over4}\right].
\eea
The first line of \Eqn{eq:ttt} is reminiscent of the fluctuation-dissipation relation \eqn{eq:FDrel2} and the second line uses the identity
\beq
\label{eq:relation}
 \left(n(2x)+{1\over2}\right)\left(n\left({x}\right)+{1\over2}\right)={1\over2}\left[\left(n\left({x}\right)+{1\over2}\right)^2+{1\over4}\right].
\eeq

As announced, the memory and noise kernels $\Sigma_\rho(t)$ and $\Sigma_F(t)$ exhibit highly non-local power law behaviors at large times. Still, as explained from rather general considerations in the previous section, there is a well-defined time regime in which the two-point correlators $F(t)$ and $\rho(t)$ can be accurately described by a Markovian Langevin dynamics with local kernels. To make the argument more precise in the present model, we now compute the explicit late-time behavior of these correlators. Again this can be done by standard contour integration techniques using Eqs. \eqn{eq:Ff} and \eqn{eq:rhof}. The calculation is actually rather similar as the previous one for self-energies with two more pairs of complex conjugate poles in the complex frequency plane coming from the retarded and advanced propagators $\tilde G_R(\omega)$ and $\tilde G_A(\omega)=\tilde G_R^*(\omega)$, located at $\pm M-i\gamma$ and $\pm M + i\gamma$ \footnote{We assume the small width limit $\gamma/M\ll1$ throughout and neglect the mass correction $\sim{\rm Re}\tilde\Sigma_R(M)$ as well as the field renormalization \eqn{eq:fieldrenorm} for simplicity.}, with 
\beq
\label{eq:gamma0}
 \gamma={i\tilde\Sigma_\rho(M)\over4M}=\frac{g^2}{16\pi M}\sqrt{1 - \frac{4m^2}{M^2}} \left[n\left({\beta M\over2}\right) + {1\over2}\right].
\eeq
The late-time behavior, $Tt\gg1$, is governed by these poles and by the branch cuts $|\omega-2m|\ge 0$ along the real frequency axis. If the poles are sufficiently far from the threshold $\omega=2m$, the pole and cut contributions are well separated. Here we assume $M \gg m$ for simplicity. We get, for the spectral function,
\beq
	\label{rhoLangevin} 
	\rho(t)    \approx  \frac{\sin{(Mt)}}{M} e^{-\gamma |t|} +\rho^{\rm pow}(t)
\eeq
and, for the equilibrium statistical one,
\beq
	\label{FLangevin}
	F^{\rm eq} (t)   \approx  \left(n\left(\beta M\right) + {1\over2}\right) \frac{\cos Mt}{M} e^{- \gamma |t|}+F^{\rm pow}(t)+F_\infty.
\eeq
In both cases, the first term on the right-hand-side is due to the poles whereas the second ones arise from the branch cuts near threshold:
\beq
\label{eq:rhopowmodel}
	\rho^{\rm pow}(t) = -\big|\tilde G_R(2m)\big|^2\sigma^{\rm pow}_\rho(t)\approx\frac{\sigma^{\rm pow}_\rho(t)}{(M^2-4m^2)^2}\qquad
\eeq
and similarly for $F_\varphi^{\rm pow}(t)$ with $\sigma^{\rm pow}_\rho(t)\to\sigma^{\rm pow}_F(t)$. Finally the statistical correlator does not decay to zero at large $t$ because of the constant $\sigma_\infty$ in \eqn{latetime}: 
\beq
\label{eq:Finfty}
	F_\infty  = \sigma_\infty\big|\tilde G_R(0)\big|^2\approx{\sigma_\infty\over M^4}.
\eeq
In Eqs. \eqn{eq:rhopowmodel}-\eqn{eq:Finfty}, we used the exact expression
\beq
 |\tilde G_R(\omega)|^{-2}=\left(M^2+{\rm Re}\tilde\Sigma_R(\omega)-\omega^2\right)^2+\frac{1}{4}\tilde\Sigma_\rho^2(\omega)\,,
\eeq
as well as the fact that at the two-particle threshold $\tilde\Sigma_\rho(2m)=0$, see \Eqn{freefieldsigmarho}.

We are now in a position to obtain a more precise estimate of the time range \eqn{eq:trange} in which a Langevin description with local memory kernel (and thus, in the present case, also a white, Markovian noise as explained in the previous section) provides a valid description in the present model. The power law contribution $\rho^{\rm pow}$ is at most of relative order $\gamma/M$ as compared to the exponential one in \eqn{rhoLangevin} for times $t\gtrsim t_\rho^L$ with, in the limit where $\gamma t^{L}_\rho\gg1$,
\beq
 		\label{time_regime0}
		{\gamma t^{L}_\rho}\approx\ln\left\{\left(\frac{m}{\gamma}\right)^{\!3/2}\!\frac{M}{m}\left(1-\frac{4m^2}{M^2}\right)^{\!5/2} \frac{n\left({\beta M\over2}\right) +{1\over2}}{n(\beta m)+{1\over2}}\right\},
\eeq
where we used \eqn{eq:relation} as well as \Eqn{eq:gamma0} to trade the coupling $g^2$ for the damping rate $\gamma$. We shall refer to $t_\rho^L$ as the Langevin time. Demanding a similar constraint on the power law contribution in \eqn{FLangevin}  leads to a slightly more constraining condition ($t^{L}_F<t^{L}_\rho$ for $M>2m$):
\beq
 	\label{time_regime1}
	{\gamma t^{L}_F}\approx{\gamma t^{L}_\rho}+\ln\left\{\frac{n\left({\beta M}\right) +{1\over2}}{n(2\beta m)+{1\over2}}\right\}.
\eeq

Let us quote the relatively simpler expressions of the Langevin time \eqn{time_regime0} in the following cases.  At low temperature, $1\ll\beta m\ll\beta M$,
\beq
		\label{time_regime3}
 		{\gamma t_\rho^{L}} \approx\ln\left[\left(\frac{m}{\gamma}\right)^{\!3/2}\!\frac{M}{m} \right];
\eeq
At intermediate temperature, $\beta m\ll1\ll\beta M$,
\beq
 		\label{time_regime4}
		{\gamma t_\rho^{L}} \approx \ln\left[\left(\frac{m}{\gamma}\right)^{\!3/2}\!\frac{M}{T} \right];
\eeq
At high temperatures, $\beta m\ll\beta M\ll1$,
\beq
 		\label{time_regime2}
		{\gamma t_\rho^{L}} \approx\ln\left(\frac{m}{\gamma}\right)^{\!3/2}.
\eeq
It is interesting to notice that the Langevin time is governed by the ratio $\gamma/m$ and not $\gamma/M$. 

We close this section by mentioning that in the case $M < 2m$, the on-shell damping rate $\gamma = 0$ such that one has formally $t_{\rho,F}^L \to \infty$ at one-loop. In this case, where dissipative processes $\varphi\to\chi\chi$ are kinematically forbiden, the pole contributions in Eqs. \eqn{rhoLangevin}-\eqn{FLangevin} are not damped and the power laws never dominates, hence $t_{\rho,F}^L \to \infty$. In fact, in this case damping arises at two-loop order. One has $\gamma\propto g^4$ and the Langevin time is parametrically reduced by a $\ln g^2\sim\ln\sqrt\gamma$ contribution.

\section{Consistency check}
\label{sec:check}

We have shown an explicit example where the relevant self-energies exhibit a highly non local behavior, but where a local Langevin desciription still provides an accurate description in the time range \eqn{time_regime1}. It is instructive, for a deeper understanding of how this comes about as well as for discussing more intricate situations where an exact solution of the non-equilibrium equations of motion is not available, to analyze this issue directly in real time, at the level of the memory integrals in Eqs. \eqn{eq:Ffourier}-\eqn{eq:GRAfourier}, in the spirit of our discussion in Subsec. \ref{sec:as}. To this aim, we shall perform the following consistency check, coming back to the general discussion of Sec. \ref{sec:exact}: We plugg the solutions of the effective local description obtained in Subsec. \ref{sec:MLd} back in the original memory integrals of Eqs. \eqn{eq:Ffourier}-\eqn{eq:GRAfourier} and check whether the power law decay of the memory kernel $\Sigma_p^\rho(t)$ is sufficient for the validity of the local approximation \eqn{eq:approxlocal}.

We begin with \Eqn{eq:GRAfourier} or, equivalently, \Eqn{eq:rhop} for the spectral function $\rho_p(t)$.
It is convenient to write the effective solution \eqn{sol:rhomarkov} as 
\beq
 \rho_p^{\rm eff}(t)=\frac{\sin\epsilon_p t}{\epsilon_p}e^{-\gamma_p|t|}=-{1\over\epsilon_p}{\rm Im}\left\{e^{-i\epsilon_pt-\gamma_p|t|}\right\},
 \label{sol:rhoeff}
\eeq
which leads, for $t>0$, to:
\beq
 \dot\rho_p^{\rm eff}(t)+\gamma_p\rho_p^{\rm eff}(t)={\rm Re}\left\{e^{-(i\epsilon_p+\gamma_p)t}\right\}.
\label {eq:trick}
\eeq
Expressing $\rho_p(t-\tau)$ in terms of $\rho_p(t)$ and $\partial_t\rho_p(t)=\dot\rho_p(t)$, we get
\beq
  \rho_p^{\rm eff}(t-\tau)=\rho_p^{\rm eff}(t){\rm Re}\left[e^{z\tau}\right]-\frac{\dot\rho_p^{\rm eff}(t)+\gamma_p\rho_p^{\rm eff}(t)}{\epsilon_p}{\rm Im}\left[e^{z\tau}\right],
  \label{eq:trick2}
\eeq
with $z=i\epsilon_p+\gamma_p$. Using this expression to evaluate the memory integral in \Eqn{sol:rhomarkov}, we obtain
\bea
&& \int_0^t d\tau\,\Sigma_p^\rho(t-\tau)\rho_p^{\rm eff}(\tau)= \int_0^t d\tau\,\Sigma_p^\rho(\tau)\rho_p^{\rm eff}(t-\tau)\nn
&&=\rho_p^{\rm eff}(t){\rm Re}\,\sigma_p^z(t)-\frac{\dot\rho_p^{\rm eff}(t)+\gamma_p\rho_p^{\rm eff}(t)}{\epsilon_p}{\rm Im}\,\sigma_p^{z}(t)\,,
\label{eq:effdamping}
\eea
where
\beq
\label{eq:integral}
 \sigma_p^z(t)=\int_0^t d\tau\,\Sigma_p^\rho(\tau)e^{z\tau}.
\eeq

We are interested in the large time behavior of the integral \eqn{eq:integral}. We already see that for too large time the power law suppression due to the memory kernel $\Sigma_p^\rho$ is not enough to overcome the exponential $e^{zt}\propto e^{\gamma_pt}$ and the integral diverges exponentially, signalling the breakdown of the effective description \eqn{sol:rhoeff}. To evaluate the late time ($\epsilon_pt\gg1$) behavior of the integral \eqn{eq:integral} more precisely, we introduce a time separation $\eta$, such that $1/\epsilon_p\ll\eta\ll1/\gamma_p$ and $\eta\ll t$, and write
\beq
 \label{eq:parameter}
 \sigma_p^{z}(t)\approx\int_0^\eta d\tau\,\Sigma_p^\rho(\tau)e^{i\epsilon_p\tau}+\int_{\eta}^t \!d\tau\,\Sigma_p^\rho(\tau)e^{z\tau}\,,
\eeq
where we neglected the exponential growth in the first term on the right-hand side. For $\epsilon_p\eta\gg1$, the upper bound contribution to the latter is suppressed by the oscillating factor and one has
\beq
 \int_0^\eta d\tau\,\Sigma_p^\rho(\tau)e^{i\epsilon_p\tau}\approx\int_0^{+\infty} d\tau\,\Sigma_p^\rho(\tau)e^{i\epsilon_p\tau}=\tilde\Sigma_p^R(\epsilon_p).
\eeq
This is the required contribution for \Eqn{eq:effdamping} to reduce to the desired local description, up to ${\cal O}(\gamma_p/\epsilon_p)$ contributions, see \Eqn{eq:localll} below.

To evaluate the second term on the right-hand side of \eqn{eq:parameter}, we use the large time, power law behavior of the memory kernel, which we parametrize as
\beq
 \Sigma_p^\rho(t)\sim \frac{a_\rho}{(\mu t)^\nu}\sigma(\mu t)\,,
 \label{eq:ansatz}
\eeq 
where $a_\rho$ is a dimensionful  amplitude parameter, typically proportional to some positive power of the relevant coupling constant, $\mu$ is a characteristic mass scale, $\sigma(x)\sim 1$ is a bounded (e.g. periodic) function and $\nu$ a given exponent \footnote{All these quantities may in principle depend on the momentum $p$. We omit the $p$ index for simplicity.}. We show in Appendix \ref{appsec:asymptotics} that, for $|zt|\to\infty$,
\beq
 \label{eq:asymptotics}
 \int_{\eta}^t \!d\tau\,\Sigma_p^\rho(\tau)e^{z\tau}\sim {e^{zt}\over z} \frac{ a_\rho}{(\mu t)^\nu}\sum_{n\ge0}\left(-\frac{\mu}{z}\right)^n\sigma^{(n)}(\mu t)\,,
\eeq
where $\sigma^{(n)}(x)=d^n\sigma(x)/dx^n$. Recalling the definitions \eqn{def:delta} and \eqn{def:gamma}, we finally obtain, for $\epsilon_pt\gg1$,  
\beq
\label{eq:lessstringent}
 {\rm Re} \,\sigma_p^{z}(t)={\rm Re}\,\tilde\Sigma_p^R(\epsilon_p)\left[1+{\cal O}\left(\frac{a_\rho}{\epsilon_p\delta\epsilon_p^2}{e^{\gamma_pt}\over(\mu t)^\nu}\right)\right]
\eeq
and
\beq
\label{eq:stringent}
 {\rm Im} \,\sigma_p^{z}(t)={\rm Im}\,\tilde\Sigma_p^R(\epsilon_p)\left[1+{\cal O}\left(\frac{a_\rho}{\gamma_p\epsilon_p^2}{e^{\gamma_pt}\over(\mu t)^\nu}\right)\right].
\eeq

As long as the last terms in brackets in the above equations can be neglected, the memory integral \eqn{eq:effdamping} has the desired form:
\beq
\label{eq:localll}
 \int_0^t d\tau\,\Sigma_p^\rho(t-\tau)\rho_p^{\rm eff}(\tau)\approx\delta\epsilon_p^2\rho_p^{\rm eff}(t)+2\gamma_p\dot\rho_p^{\rm eff}(t)\,,
\eeq
up to ${\cal O}(\gamma_p/\epsilon_p)$ corrections, with
\beq
\label{eq:localllparam}
 \delta\epsilon_p^2={\rm Re}\,\tilde\Sigma^R_p(\epsilon_p)\,,\qquad \gamma_p=-\frac{{\rm Im}\tilde\Sigma^R_p(\epsilon_p)}{2\epsilon_p}\,.
\eeq
The ansatz \eqn{sol:rhoeff} is indeed a solution of \Eqn{eq:rhop} with \eqn{eq:localll} and the local approximation is thus consistent within a finite time range governed by the terms in brackets in Eqs. \eqn{eq:lessstringent}-\eqn{eq:stringent}. For the local description -- which in the present case gives rise to an exponentially decaying solution -- to make sense, the latter should be  larger than $\gamma_p^{-1}$ which typically requires a weak coupling situation. The ratio $a_\rho/\epsilon_p\delta\epsilon_p^2$ is parametrically of order one in the relevant coupling, whereas the ratio $a_\rho/\gamma_p\epsilon_p^2$ may either be of order one in the case where damping is allowed at lowest order in perturbation theory, or be enhanced by inverse powers of the coupling if damping is only possible at higher orders. Thus the most stringent restriction comes from \Eqn{eq:stringent} and we check that the Langevin time, for which the local description makes sense, i.e.
\beq
\label{eq:ltimeagain}
 \gamma_pt^L\approx\ln\left\{\left(\frac{\mu}{\gamma_p}\right)^{\!\nu}\frac{\gamma_p\epsilon_p^2}{a_\rho}\right\},
\eeq
 agrees with the analysis of the previous sections, see \Eqn{eq:trange}, or \Eqn{time_regime0} for the model discussed in  Sec. \ref{sec:explicit}.

 A similar -- although more lengthy -- analysis can be performed for the $F$-equation with the same conclusion, namely that, within the time range \eqn{eq:ltimeagain},
\beq
 \int_0^\infty \!\!d\tau\,\Sigma_p^\rho(\tau)F_p^{\rm eff}(t-\tau,t')\!\approx\!\delta\epsilon_p^2F_p^{\rm eff}(t,t')+2\gamma_p\partial_t F_p^{\rm eff}(t,t'),
\eeq
up to corrections of relative order ${\cal O}\left(\gamma_p/\epsilon_p\right)$, where $F_p^{\rm eff}(t,t')$ is given by \Eqn{eq:fullsolfourier} with $F_p^K(t.t')$ given by \eqn{eq:exprel} and $F_{p}^{\rm eq}(t-t')$ replaced by $F_{{\rm eff},p}^{\rm eq}(t-t')$ in \eqn{eq:Feqeff}. This completes our consistency check for the full non-equilibrium solution.

\section{Application: decoherence, entropy production and thermalization}

We end this paper with a simple application of the previous considerations to the physics of quantum decoherence, entropy production and thermalization in quantum field theory in the context of the so-called incomplete description picture, recently advocated in \cite{Campo:2008ju,Giraud:2009tn,Koksma:2009wa,Koksma:2010dt,Koksma:2011dy,Prokopec:2012xv,Koksma:2010zi,Gautier:2011fx,Prokopec:2012xv}. This is based on the observation that one's ability to measure the state of a quantum field is limited because of a restricted access to the infinite tower of $n$-point correlation functions: One typically has only access to low order correlators, most often to the subset of one- and two-point functions. The lack of knowledge of higher-order correlators may result, from the point of view of the observer, in effective loss of quantum purity and/or coherence and associated entropy production, even to effective thermalization \cite{Giraud:2009tn,Koksma:2009wa}.

Let us consider the case where only the subset of independent equal-time two-point functions
\bea
 F_p(t)&=&\left.F_p(t,t')\right|_{t'=t}\,,\nn
 R_p(t)&=&\left.\partial_tF_p(t,t')\right|_{t'=t}\,,\\
 K_p(t)&=&\left.\partial_t\partial_{t'}F_p(t,t')\right|_{t'=t} \nonumber
\eea
of a given field mode is measured at each time. From this knowledge one can reconstruct the least biased quantum state compatible with the measured correlators and infer its quantum properties. The corresponding density operator is a Gaussian in the field operators, characterized by the intrinsic -- canonically invariant -- occupation number \cite{Campo:2008ju,Giraud:2009tn,Koksma:2009wa}
\beq
\label{eq:occu}
 n_p(t)+{1\over2}=\sqrt{F_p(t)K_p(t)-R_p^2(t)}\,.
\eeq
The latter measures the quantum purity of the Gaussian state \cite{Giraud:2009tn}:
\beq
\label{eq:purity}
 P_p(t)=\frac{1}{2n_p(t)+1}\,,
\eeq
which is equal to its maximum value $1$ ($n_p=0$) for a pure state. The occupation number \eqn{eq:occu} also measures the phase space area covered by the Gaussian state in the Wigner representation and can be related to the Gaussian entropy \cite{Koksma:2009wa}
\beq
\label{eq:entropy}
 s_p(t)=[n_p(t)+1]\ln[n_p(t)+1]-n_p(t)\ln n_p(t)\,,
\eeq
which measures the amount of missing information in the subset of measured correlators, that is the amount of information stored in unmeasured higher-order correlators.

Non-intrinsic -- i.e. basis dependent -- properties of the inferred Gaussian quantum state can be characterized by introducing another occupation number (note that $0\le n_p(t)\le\bar n_p(t)$)
\beq
 \bar n_p(t)+{1\over2}=\frac{K_p(t)+\epsilon_p^2F_p(t)}{2\epsilon_p}\,.
\eeq
For instance, the squeezing, or coherence parameter
\beq
\label{eq:decoparam}
 g_p(t)=\sqrt{1-\left(\frac{n_p(t)+{1/2}}{\bar n_p(t)+{1/2}}\right)^2}
\eeq
measures the degree of quantum entanglement/coherence in the basis of semi-classical coherent states. A state with $g_p\to1$ exhibits non-trivial correlations between macroscopically distant semi-classical states and thus a high degree of quantum coherence (in this basis) \cite{Campo:2008ju,Giraud:2009tn}.

Now to the dynamics. We consider the simple situation of Sec. \ref{sec:exact} where the field of interest is weakly coupled to a thermal bath with negligible backreaction. The exact solution of the nonequilibrium dynamics is given by Eqs. \eqn{eq:fullsolfourier}-\eqn{eq:CImom}, with \eqn{eq:rhopow}. In the small width limit, the Langevin time \eqn{eq:trange} is large and the power law contribution in Eqs. \eqn{eq:Fpow} and \eqn{eq:exprel}-\eqn{CI_again} can be neglected, resulting in a Markovian dynamics as explained in Subsec. \ref{sec:MLd}. Reshuffling the various terms (hence the bar on $\bar F_p^0$ below), one can write, for $t,t'>0$, 
\bea
 F_p(t,t')=&&F_p^{\rm eq}(t-t')\left(1-e^{-\gamma_p\left[\left(t+t'\right) - \left|t-t'\right|\right]}\right) \nn
 && + \bar F_p^0(t,t')e^{-\gamma_p(t+t')}\,,
\eea
where 
\beq
 \bar F^0_p(t,t')\!=\! \bar a^+_p\!\cos\epsilon_p(t-t')+ \bar a^-_p\!\cos\epsilon_p(t+t')+ \bar a^0_p\sin\epsilon_p(t+t'),
\eeq
with, neglecting ${\cal O}(\gamma_p/\epsilon_p)$ corrections and setting $Z_p\approx1$,
\beq
 \bar a^\pm_p =  \frac{1}{2}\left(F_p(0)\pm \frac{K_p(0)}{\epsilon_p^2}\right) \,,\quad\bar a^0_p =  \frac{R_p(0)}{\epsilon_p} \,.
\label{CI_againbis}
\eeq

In the Breit-Wigner -- or Markov -- approximation considered here, the equal-time equilibrium two-point correlators are characterized by the on-shell equilibrium occupation number $n_p^{\rm eq}=n(\beta\epsilon_p)$:
\beq
 F_p^{\rm eq}(0)={1\over\epsilon_p}\left(n_p^{\rm eq}+{1\over2}\right)\,,
\eeq
with $R_p^{\rm eq}(0)=0$ and $K_p^{\rm eq}(0)=\epsilon_p^2F_p^{\rm eq}(0)$. It is an simple exercice to compute the various quantities \eqn{eq:occu}-\eqn{eq:decoparam}. One obtains
\beq
 n_p(t)+{1\over2}=\sqrt{\left[N_p^{\rm eq}(t)\right]^2+2N_p^{\rm eq}(t)\bar N_p^{0}(t)+\left[N_p^{0}(t)\right]^2}\,,
\eeq
where we defined
\beq
 N_p^{\rm eq}(t)=\left(n_p^{\rm eq}+{1\over2}\right)\left(1-e^{-2\gamma_pt}\right)\,,
\eeq
\beq
 \bar N_p^0(t)=\left(\bar n_p(0)+{1\over2}\right)e^{-2\gamma_pt}
\eeq
and
\beq
 N_p^0(t)=\left(n_p(0)+{1\over2}\right)e^{-2\gamma_pt}\,.
\eeq
The occupation number $\bar n_p(t)$ and the decoherence parameter $g_p(t)$ have remarkably simple expressions:
\beq
 \bar n_p(t)=n_p^{\rm eq}\left(1-e^{-2\gamma_pt}\right)+\bar n_p(0)e^{-2\gamma_pt}
\eeq
and
\beq
 g_p(t)=g_p(0)\,\frac{\bar n_p(0)+1/2}{\bar n_p(t)+1/2}e^{-2\gamma_pt}\,.
\eeq
\begin{figure}[t!]  
\hspace{-3.cm} \epsfig{file=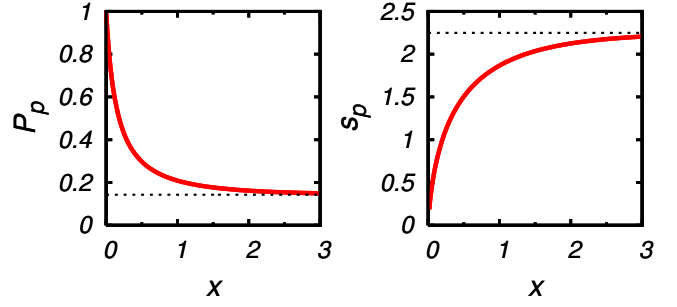,width=6.cm}
 \caption{\label{fig:purity-entropy-1} 
 The Gaussian purity \eqn{eq:purity} (left) and entropy \eqn{eq:entropy} (right) as functions of $x=2\gamma_pt$ for the case of an initial vacuum-like -- i.e. pure but with no long range coherence in the basis of coherent states -- state: $n_p(0)=\bar n_p(0)=0$. The equilibrium occupation number, which determines the asymptotic values of both the purity and entropy (dashed lines), is $n_p^{\rm eq}=3$.}
\end{figure}
Note that, as expected, an initially incoherent, or thermal-like state with $g_p(0)=0$ or, equivalently $n_p(0)=\bar n_p(0)$, remains so: $g_p(t)=0$, or $n_p(t)=\bar n_p(t)$. A particular case is $n_p(0)=\bar n_p(0)=0$, which corresponds to a pure, vacuum-like state. Preparing such a state at time $t=0$, one observes an effective loss of quantum purity due to the incomplete knowledge of higher order correlators. Indeed, in that case, the Gaussian purity \eqn{eq:purity} reads
\beq
 P_p(t)=\frac{P_p^{\rm eq}}{1-(1-P_p^{\rm eq})e^{-2\gamma_pt}}\,,
\eeq
where $P_p^{\rm eq}=1/(2n_p^{\rm eq}+1)$. This loss of purity naturally corresponds to a loss of information and thus to a growth of entropy. \Fig{fig:purity-entropy-1} shows the purity \eqn{eq:purity} and the entropy \eqn{eq:entropy} as a function of $x=2\gamma_pt$ for this case.

\begin{figure}[t!]  
\hspace{-3.cm} \epsfig{file=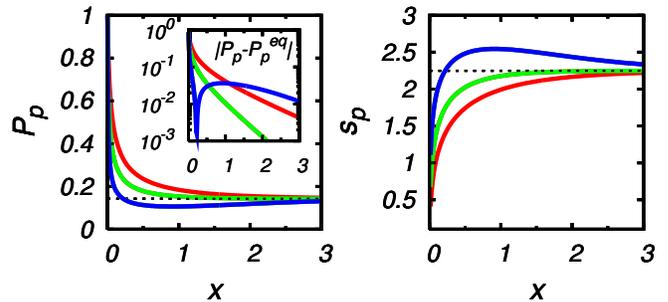,width=6.cm}
 \caption{\label{fig:purity-entropy-2} The Gaussian purity \eqn{eq:purity} (left) and entropy \eqn{eq:entropy} (right) as functions of $x=2\gamma_pt$ for the case of an initial squeezed vacuum-like -- i.e. pure and highly coherent -- state: $n_p(0)=0$ and $\bar n_p(0)=1,3,10$ (top to bottom in the left panel and bottom to top in the right one), corresponding to initial coherence factors $1-g_p(0)=5\times10^{-2},10^{-2},10^{-3}$ respectively. The equilibrium occupation number, which determines the asymptotic values of both the purity and entropy (dashed lines), is $n_p^{\rm eq}=3$. The insert shows the approach to the equilibrium value $P_p^{\rm eq}$ in logarithmic scale. One sees that the relaxation rate is twice faster in the case $\bar n_p(0)=n_p^{\rm eq}$. The entropy, being one to one related to purity, exhibits a similar behavior.}
\end{figure}

\begin{figure}[t!]  
\hspace{-3.cm} \epsfig{file=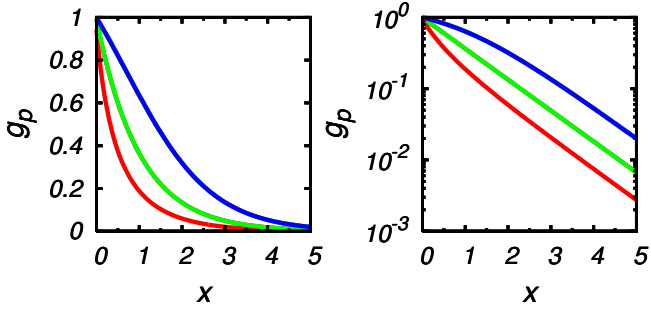,width=6.cm}
 \caption{\label{fig:deco} The coherence parameter \eqn{eq:decoparam} as a function of $x=2\gamma_pt$ in linear (left) and logarithmic (right) scales for the same parameters as in \Fig{fig:purity-entropy-2}, here $\bar n_p(0)=1,3,10$ bottom to top. The case $\bar n_p(0)=n_p^{\rm eq}$ leads to a purely exponential decay.}
\end{figure}

Let us now consider the case of an initial pure state, $n_p(0)=0$, with a high degree of quantum coherence $g_p(0)\to1$, i.e. $\bar n_p(0)\gtrsim1$, which corresponds to a squeezed vacuum-like state. Again because information about the state of the systems spreads towards higher order correlators, one observes apparent loss of quantum purity and associated entropy growth as well as apparent loss of quantum coherence. This is illustrated in Figs. \ref{fig:purity-entropy-2} and \ref{fig:deco} for $n_p^{\rm eq}=3$ and $\bar n_p(0)=1,3,10$.

The case $\bar n_p(0)=n_p^{\rm eq}$, which implies that $\bar n_p(t)=n_p^{\rm eq}$ is particular. For instance, one gets
\beq
 n_p(t)+{1\over2}= \left(n_p^{\rm eq}+{1\over2}\right)\sqrt{1-\frac{n_p^{\rm eq}(n_p^{\rm eq}+1)}{n_p^{\rm eq}+1/2}e^{-4\gamma_pt}}\,,
\eeq
and thus the Gaussian purity and entropy approach their late time equilibrium value at a rate $4\gamma_p$, twice the rate of the case $\bar n_p(0)\neq n_p^{\rm eq}$. This is illustrated in the insert of \Fig{fig:purity-entropy-2}. Also the coherence factor has a purely exponential decay law
\beq
 g_p(t)=g_p(0)e^{-2\gamma_pt}\,,
\eeq
shown in \Fig{fig:deco}.

For the case $\bar n_p(0)>n_p^{\rm eq}$ purity rapidly falls off and undershoots -- entropy correspondingly overshoots -- its equilibrium value before exponentially approaching it at a rate $2\gamma_p$. The time $t_{\rm ext}$ at which an extremum is reached (minimum of $P_p$ or maximum of $s_p$) decreases when $\bar n_p(0)$ is increased but remains bounded: $2\gamma_pt_{\rm ext}>\ln 2$. Thus the time scale of purity loss/entropy growth is always set by $\gamma_p^{-1}$. Observe finally that cases with $\bar n_p(0)>n_p^{\rm eq}$ have faster loss of purity or growth of entropy than cases with $\bar n_p(0)<n_p^{\rm eq}$, but they have slower loss of coherence.

\section{Summary and Conclusion}

To summarize, we have seen that the dynamics of non-equilibrium quantum fields can always be formally written as an equivalent, non-local, non-linear Langevin process with self-consistently determined memory and noise kernels. A drastic simplification is achieved when the memory integral can be approximated by a local damping term. It is often assumed that this requires the existence of a finite, short memory time scale beyond which the memory kernel essentially vanishes. The typical situation in QFT is, however, that the latter exhibit a power law behavior at large times. Still, such a power law decay, if not enough to completely localize the memory integrals, is sufficient to make the dynamics essentially local in a finite time interval -- before late-time power laws come to dominate -- at least if the associated local damping rate is weak enough. 

We have illustrated the above points in a simple example with non-equilibrium degrees of freedom interacting with a stationary thermal bath with negligible back-reaction, where an analytical solution of the QFT problem is possible. We believe our analysis clarifies the validity of local descriptions of non-equilibrium field dynamics, as well as of related Breit-Wigner motivated approximations, employed in a wide range of problems, see e.g. \cite{Herranen:2008di,Herranen:2008hu,Fidler:2011yq,Garbrecht:2011aw,Drewes:2012qw}. Finally, we have provided a general method to work out the local limit directly at level of the equations of motion, which is useful when a complete solution is not known, for instance in non-stationary situations, or when back-reaction is important -- in which case the damping rate as well as the noise kernel may depend self-consistently on the non-equilibrium correlators. 
 
As an illustration, we have applied the above ideas to the description of quantum decoherence and entropy production in QFT in a simple situation with a stationary thermal environment along the lines of Refs \cite{Giraud:2009tn,Koksma:2009wa}. Interesting further applications of the methods developed here include, for instance, the study of decoherence of primordial fluctuations in inflationary scenarios \cite{Campo:2008ju}, or the discussion of local damping descriptions in expanding geometries, which underly e.g. warm inflationary scenarios \cite{Berera:2004kc,Aarts:2007qu} or the standard perturbative description of reheating in the early Universe \cite{Allahverdi:2010xz}.

\section*{Acknowledgements}
We acknowledge useful discussions with M.Drewes, M. Garny and U. Reinosa.

\appendix

\section{Calculation of $\sigma_\infty=\Sigma_{p=0}^{F}(t\to\infty)$}
\label{appsec:kappa}

From Eqs. \eqn{freefield}-\eqn{eq:sigmapt} one easily obtains, after some manipulations,
\beq
\sigma_\infty=-\frac{g^2}{8\pi}\int_{2m}^{+\infty}\!\!dx\sqrt{1-{4m^2\over x^2}}n\left({\beta x\over2}\right)\left[n\left({\beta  x\over2}\right)+1\right].
\eeq
Using $n'(x)=-n(x)(n(x)+1)$, one obtains the alternative, simpler expressions
\bea
 \sigma_\infty&=&-\frac{g^2}{4\pi\beta}\int_1^{+\infty}\!\!dx\frac{n(\beta m x)}{x^2\sqrt{x^2-1}}\nn
 &=&-\frac{g^2}{4\pi\beta}\int_0^1\!dx\,\frac{x}{\sqrt{1-x^2}}\,n\!\left({\beta m\over x}\right).
\eea
For instance, one obtains, at low temperature $\beta m\gg1$
\beq
\sigma_\infty=-\frac{g^2e^{-\beta m}}{8\pi\beta}\sqrt{\frac{2\pi}{\beta m}}\left[1+{\cal O}\left(1/\beta m\right)\right],
\eeq
whereas in the high temperature limit $\beta m\ll1$
\beq
\sigma_\infty=-\frac{g^2}{16\beta^2 m}\left[1+{\cal O}\left(\beta m\right)\right].
\eeq

It is interesting to see how this constant arises in the standard calculation in frequency space, where one first computes $\tilde\Sigma_{p=0}^\rho(\omega)$ and deduce $\tilde\Sigma_{p=0}^F(\omega)$ by using the fluctuation-dissipation relation \eqn{eq:FDrel2}. This seems to leave no space for the required $\delta(\omega)$ contribution in $\tilde\Sigma_{p=0}^F(\omega)$. In fact, a careful evaluation of $\tilde\Sigma_{p=0}^\rho(\omega)$ reveals a contribution 
\beq
 \left[(n(\beta k_0)+{1\over2}+n(\beta (\omega-k_0))+{1\over2}\right] \delta(\omega)\,,
\eeq
where $k_0$ is a frequency variable to be integrated over. This contribution can be safely put to zero in $\Sigma_{p=0}^\rho(\omega)$, owing to the property $n(-x)+1/2=-(n(x)+1/2)$. However, it leads to a non-vanishing contribution when multiplied by $n(\beta\omega)+1/2$ to compute $\Sigma_{p=0}^F(\omega)$ as demanded by the fluctuation dissipation relation. Indeed, using the identity
\bea
 \left[(n(\beta k_0)+{1\over2}+n(\beta (\omega-k_0))+{1\over2}\right] \left(n(\beta\omega)+{1\over2}\right)\nn
=\left(n(\beta k_0)+{1\over2}\right)\left(n(\beta (\omega-k_0))+{1\over2}\right)+{1\over4}\,,\nn
\eea
which is easily demonstrated by using $2n(x)+1=\tanh^{-1}(x/2)$ and making use of standard trigonometric relations, one obtains a non-vanishing term in $\Sigma_{p=0}^F(\omega)$:
\beq
 \left[-\left(n(\beta k_0)+{1\over2}\right)^2+{1\over4}\right]\delta(\omega).
\eeq
One easily checks that this term is also obtained by a direct calculation of  $\Sigma_{p=0}^F(\omega)$, without using the fluctuation dissipation relation. Finally, one checks that the final coefficient of $\delta(\omega)$ is precisely the constant $\sigma_\infty$ obtained by the direct calculation in the time-momentum representation described above. 

\section{Asymptotic behavior of $\sigma^z_p(t)$}
\label{appsec:asymptotics}

This section is devoted to the asymptotic behavior of  $\int_{\eta}^t \!d\tau\,\Sigma_p^\rho(\tau)e^{z\tau}$ in \Eqn{eq:parameter} as $|zt|\to\infty$. Integrating $n$-times  by part reads:
\bea
 \int_{\eta}^t \!d\tau\,\Sigma_p^\rho(\tau)e^{z\tau} &=& \sum_{k = 0}^{n-1}\frac{(-1)^k}{z^{k+1}}\left\{e^{zt}s_k(t)-e^{z\eta}s_k(\eta)\right\} \nn
 &&+ (-1)^n \int_{\eta}^t \!d\tau\,s_n(\tau)\frac{e^{z\tau}}{z^n}\,,
\eea
where $ s_n(t)=\frac{d^n}{dt^n}\Sigma_p^\rho(t)$. We recall that in the generic case considered here $\Sigma_p^\rho(t)$ is a power law and therefore $\mathcal{C}^\infty$. Thus for any finite interval $[\eta, t]$, $\int_{\eta}^t \!d\tau\,s_n(\tau)\frac{e^{z\tau}}{z^n} \to 0$ as $n \to \infty$. This allows us perform an infinite amount of integration by parts which gives 
\beq
 \int_{\eta}^t \!d\tau\,\Sigma_p^\rho(\tau)e^{z\tau} = \sum_{n \ge 0}\frac{(-1)^n}{z^{n+1}}\left\{e^{zt}s_n(t)-e^{z\eta}s_n(\eta)\right\}\,.
\eeq

Moreover, since ${\rm Re }\, z>0$, the contribution from the lower boundary is exponentially suppressed. Using the generic ansatz \eqn{eq:ansatz} and the fact that $\sigma$ is bounded, we get: 
\beq
\int_{\eta}^t \!d\tau\,\Sigma_p^\rho(\tau)e^{z\tau}\sim \frac{a_p e^{zt}}{z(\mu_p t)^\nu}{\cal F}\left(\mu_p t,\frac{\mu_p}{z}\right)\,,
\eeq
where 
\beq
 {\cal F}\left(x,y\right)=\sum_{n\ge0}(-y)^n\sigma^{(n)}(x)\left[1+{\cal O}(1/x)\right]\,.
\eeq

The above expression simplifies when the function $\sigma(x)$ is periodic and thus satisfies $\sigma^{(2p)}(x)=(-1)^p \sigma(x)$ and $\sigma^{(2p+1)}(x)=(-1)^p \sigma'(x)$. Splitting the sum into odd and even terms, one gets, for $|y|<1$,
\beq
 {\cal F}\left(x,y\right)=\frac{\sigma(x)-y\sigma'(x)}{1+y^2} \left[1+{\cal O}(1/x)\right]\,.
\eeq
As  $|\mu_p^2/z^2|=\mu_p^2/(\epsilon_p^2+\gamma_p^2)<1$ we get 

\beq
 \int_{\eta}^t \!d\tau\,\Sigma_p^\rho(\tau)e^{z\tau}\sim \frac{e^{zt}}{(\mu t)^\nu}\frac{z\sigma(\mu t)-\mu\sigma'(\mu t)}{z^2+\mu^2}\,.
\eeq

In the case where $\mu\ll\epsilon_p$ this further simplifies to
\beq
 \int_{\eta}^t \!d\tau\,\Sigma_p^\rho(\tau)e^{z\tau}\sim \frac{e^{zt}\sigma(\mu t)}{i\epsilon_p(\mu t)^\nu}=\frac{e^{zt}}{i\epsilon_p}\Sigma_p^\rho(t)\,.
\eeq

\bibliographystyle{h_physrev}
\bibliography{biblio}
\nocite{}

\end{document}